\RequirePackage{fix-cm}
\documentclass[twocolumn]{svjour3}          
\smartqed  
\usepackage{graphicx}
\usepackage[noend]{algpseudocode}
\usepackage{algorithmicx,algorithm}
\usepackage{amsmath}
\usepackage[misc]{ifsym}
\usepackage[authoryear]{natbib}
\usepackage{color}

\usepackage{cite}
\usepackage{natbib}
\setcitestyle{numbers,square}

\begin{document}

\title{A Quantum System Control Method Based on Enhanced Reinforcement Learning
}


\author{Wenjie Liu\textsuperscript{1,2}         \and
        Bosi Wang\textsuperscript{1} \and
        Jihao Fan\textsuperscript{3} \and
        Yebo Ge\textsuperscript{1}\and
        Mohammed Zidan\textsuperscript{4}
}

\institute{%
    \begin{itemize}
      \item[\textsuperscript{\Letter}] {Wenjie Liu} \\
            {wenjiel@163.com}
       \item[]{Bosi Wang} \\
            {bosi@nuist.edu.cn}
      \at
      \item[\textsuperscript{1}] School of Computer and Software, Nanjing University of Information Science and Technology, Nanjing 210044, China
      \item[\textsuperscript{2}] Engineering Research Center of Digital Forensics Ministry of Education, Nanjing 210044, China
      \item[\textsuperscript{3}] School of Electronic and Optical Engineering, Nanjing University of Science and Technology, Nanjing 210094, China
      \item[\textsuperscript{4}] Hurghada Faculty of Computers and Artificial Intelligence, South Valley University, Egypt
    \end{itemize}
}

\date{Received: date / Accepted: date}
\maketitle

\begin{abstract}
Traditional quantum system control methods often face different constraints, and are easy to cause both leakage and stochastic control errors under the condition of limited resources. Reinforcement learning has been proved as an efficient way to complete the quantum system control task. To learn a satisfactory control strategy under the condition of limited resources, a quantum system control method based on enhanced reinforcement learning (QSC-ERL) is proposed. The states and actions in reinforcement learning are mapped to quantum states and control operations in quantum systems. By using a new enhanced neural networks, reinforcement learning can quickly achieve the maximization of long-term cumulative rewards, and a quantum state can be evolved accurately from an initial state to a target state. According to the number of candidate unitary operations, the three-switch control is used for simulation experiments. Compared with other methods, the QSC-ERL achieves close to 1 fidelity learning control of quantum systems, and takes fewer episodes to quantum state evolution under the condition of limited resources.
\end{abstract}

\section{Introduction}
\label{intro}
Quantum system control is one of the keys to the development of quantum information technology, which has been applied in many fields, such as  multi-photon interference measurement (Vedaie et al. \citeyear{vedaie2018reinforcement}), quantum error correction (Fosel et al. \citeyear{fosel2018reinforcement}), quantum states preparation (Bukov et al. \citeyear{bukov2018reinforcement}). Since most quantum systems cannot meet the two constraint conditions, one is strong regular free Hamiltonian, the other is that interaction Hamiltonian are fully connected (Meng and Cong \citeyear{meng2022control}), it is difficult to implement active manipulation or control. In order to manipulate quantum systems to a ideal performance, different control methods have been developed (Patsch  et al. \citeyear{patsch2020simulation}; An et al. \citeyear{an2021quantum}; Torosov et al. \citeyear{torosov2021coherent}).  For a quantum system with limited control resources, it is a challenge to effectively and accurately control quantum states evolution under perturbation.

Traditional learning algorithms (such as gradient algorithms (Chakrabarti and Rabitz \citeyear{chakrabarti2007quantum}; Roslund and Rabitz \citeyear{roslund2009gradient}), genetic algorithms (Tsubouchi and Momose \citeyear{tsubouchi2008rovibrational}) have shown excellent control effects under specific experimental environment. But in practical, the quantum system to be manipulated usually has different restrictions. There is a class of quantum system control problem with limited control resources. In this case, the gradient algorithms are not suitable for solving the above problems, and the genetic algorithms need a lot of experimental data to optimize the control performance that complicates the resolution of the problem.

With the advent of quantum information technology and the upsurge of machine learning (Abualigah et al. \citeyear{ABUALIGAH2021113609}; Abualigah et al. \citeyear{ABUALIGAH2021107250}; Abualigah et al. \citeyear{2021RSA}; Abualigah et al. \citeyear{2021IoD}), many researchers have found that machine learning can effectively help to find the optimal strategy to solve the control problem of quantum systems (Chunlin et al. \citeyear{chunlin2012hybrid}; Chen et al. \citeyear{chen2013fidelity}; Palittapongarnpim et al. \citeyear{palittapongarnpim2017robustness}). In particular, studies on quantum system control based on reinforcement learning have been increasing gradually. Reinforcement learning (Fang et al. \citeyear{fang2020survey}) interacts with the environment in the form of rewards and punishments. Vedaie et al. (\citeyear{vedaie2018reinforcement}) applied reinforcement learning to realize multi-photon interference measurement. Cardenas-Lopez et al. (\citeyear{cardenas2018multiqubit}) proposed a protocol for quantum reinforcement learning, which does not require coherent feedback during the learning process and can be implemented in a variety of quantum systems. Fosel et al. (\citeyear{fosel2018reinforcement}) showed how a network-based ``agent" can discover a complete quantum error correction method to protect qubits from noise. In addition, Bukov et al. (\citeyear{bukov2018reinforcement}) used reinforcement learning to prepare the desired quantum states. They also successfully used Q-learning (Watkins et al. \citeyear{watkins1992q}) to control quantum systems (Bukov \citeyear{bukov2018reinforcement2}). Yu et al. (\citeyear{yu2019reconstruction}) used quantum reinforcement learning to make a qubit ``agent" adapt to the unknown quantum system ``environment" to achieve maximum overlap. Niu et al. (\citeyear{niu2019universal}) used deep reinforcement learning and proposed a quantum control framework for fast and high-fidelity quantum gate control optimization. Zhang et al. (\citeyear{zhang2019does}) successfully used reinforcement learning algorithm to solve a class of quantum state control problems, and made a theoretical analysis. However, the above methods have high requirements on hardware resources in practical and are not effective for solving a class of resource-constrained quantum system control problems.

In order to complete the evolution of quantum states quickly and efficiently under the condition of insufficient hardware conditions and limited numbers and types of unitary operations that can be used, a quantum system control method based on enhanced reinforcement learning (QSC-ERL) is proposed. The quantum system control problem under the condition of limited resources is modeled using reinforcement learning algorithm. By using a proposed enhanced neural networks, reinforcement learning can more quickly achieve the maximization of long-term cumulative rewards, and a quantum state can be evolved accurately from the initial state to the target state. The simulation experiment is implemented by Python programming language and Linalg tool library. The result shows that compared with other methods, the QSC-ERL can achieve high fidelity learning control of quantum systems, and takes fewer episodes to achieve quantum state evolution under the condition of limited resources.

The main contributions of this paper are: (1) Various reinforcement learning algorithms are used for validate the effectiveness and generality of quantum system control methods based on reinforcement learning. (2) A quantum system control method based on enhanced reinforcement learning (QSC-ERL) is proposed to efficiently solve the control problem of quantum systems with limited control resources.

The rest of this paper is structured as follows. In
Sec. II, we briefly overview the preliminaries about quantum system control and reinforcement learning. In Sec. III, we model the quantum system control problem and present our novel method. In Sec. IV and V, we respectively show the results of simulation experiments and draw our conclusions.

\section{Preliminaries}
\subsection{Learning control of quantum systems}
Learning control methods are powerful for solving quantum system control problems (Ma and Chen \citeyear{ma2020several}). The learning methods are often optimized by multiple iterations to realize the evolution of qubits from an initial state to the desired target state. In this paper, the task of quantum system control is set as the quantum pure state transition control problem of n-order quantum system. For the free Hamiltonian $H_0$ of n-order quantum system, its eigenstate can be defined as $D=\left\{\left|\phi_{i}\right\rangle\right\}_{i=1}^{N}$. The quantum state to be evolved $\left|\psi_{(t)}\right\rangle$ of a controlled system can be extended according to the eigenstates in set $D$: \begin{equation}
\left|\psi_{(t)}\right\rangle=\sum_{i=1}^{N} c_{i}(t)\left|\psi_{i}\right\rangle,
\end{equation}
where the complex number $c_{i}(t)$ satisfies $\sum_{i=1}^{N}\left|c_{i}(t)\right|^{2}=1$.

In order to achieve the active control of the quantum system, the control Hamiltonian $H_c$ is introduced into the control $u(t) \in L^{2}(\mathbf{R})$, which is independent of time and interacts with the quantum system. The $\left|\psi_{(t=0)}\right\rangle$ can be redefined as $\left|\psi_0\right\rangle$. The $ C(t)=\left(C_{i}(t)\right)_{i=1}^{N}$ evolves according to the Schr\"{o}dinger equation: \begin{equation}
\left\{\begin{array}{l}
\iota \hbar \dot{C}(t)=[A+u(t) B] C(t) \\
C(t=0)=C_{0}
\end{array}\right.,
\end{equation}
where $\iota=\sqrt{-1}$, $\quad C_{0}=\left(c_{0 i}\right)_{i=1}^{N}$, $\quad c_{0 i}=\left\langle\varphi_{i} \mid \psi_{0}\right\rangle$, $\quad \sum_{i=1}^{N}\left|c_{0 i}\right|^{2}=1$, $\hbar$ is the reduced Planck constant, and the matrices $A$ and $B$ correspond to the free Hamiltonian $H_0$ and the controlled Hamiltonian $H_c$ of the quantum system respectively. $U_{\left(t_{1} \rightarrow t_{2}\right)}$ represents an unitary operation for any state $\left|\psi_{\left(t_{1}\right)}\right\rangle$ of the quantum system. The $\left|\psi_{\left(t_{2}\right)}\right\rangle=U_{\left(t_{1} \rightarrow t_{2}\right)}\left|\psi_{\left(t_{1}\right)}\right\rangle$ of the quantum system is that the quantum state $\left|\psi_{\left(t_{1}\right)}\right\rangle$ evolves from time $t=t_1$ to time $t=t_2$. In addition, $U_{\left(t_{1} \rightarrow t_{2}\right)}$ can also be defined as $U_{(t)}$, $t \in\left[t_{1}, t_{2}\right]$.

In fact, if the quantum systems evolve freely without control resources limited, it can also arrive at the target state from an initial state. However, there are two unfavorable problems in this way of free evolution control: One is that it is difficult to satisfy the conditions in practice, and will waste a lot of control resources to evolve from an initial state to the desired target state. The other is that free evolutionary control has no certain control law, and is unable to be determined when the quantum system reaches the target state. Our study mainly aims at solving a class of control resource-limited quantum system control problem.

\subsection{Quantum control landscapes}
The quantum control landscapes (Chakrabarti and Rabitz \citeyear{chakrabarti2007quantum}) has provided a theoretical basis for analyzing the learning control problem of quantum systems, which can be defined as the mapping between the control Hamiltonian and the correlation value of the control performance function. The task of quantum system control can be defined as a problem of maximizing the target performance function. In other words, it can be transformed into a problem of maximizing the state transition probability from the initial state to the desired target state. For the state transition control problem, the quantum control transition can be defined as 
\begin{equation}
\begin{aligned}
J(u)=
&tr(U_{(\varepsilon, T)}|\psi_{initial}\rangle\\
&\langle\psi_{initial}|U_{(\varepsilon, T)}^{\dagger}| \psi_{target}\rangle\langle\psi_{target}|)\label{equ:3},
\end{aligned}
\end{equation}
where $tr(\cdot)$ is the trace operation, $U^{\dagger}$ is the ad-joint of $U$, $|\psi_{initial}\rangle$ is the initial quantum state, $\left|\psi_{target}\right\rangle$ is the desired target quantum state.

In this paper, it is assumed that the control set $\{ u_{j}, j=1,2, \ldots, m\}$ allowed to operate in a controlled quantum system can be given in advance, where each control $u_{j}$ corresponds to an unitary operation $U_{j}$. The goal of learning control is to evolve control from the initial state $|\psi_{initial}\rangle$ to the desired target state $|\psi_{target}\rangle$, and learn a global optimal control sequence $u^*$:
\begin{equation} 
u^{*}=\mathop{\arg\max}\limits_{u} J(u)\label{equ:4}.
\end{equation}

\subsection{Reinforcement learning}
Reinforcement learning (Fang et al. \citeyear{fang2020survey}) is described by Markov Decision Process (MDP), which is usually defined by the quadruple $\langle S, A, P, R\rangle$. The $S$ is the set of states, $A$ is the set of actions, and the state $s \in S$, the action $a \in A .$ The state transition function $P\left(s, a, s^{\prime}\right)$ represents the probability of state transition. The $R\left(s, a, s^{\prime}\right)$ represents the reward value function. $P\left(s, a, s^{\prime}\right)$ and $R\left(s, a, s^{\prime}\right)$ only depend on the current state $s$ and action $a$ that have nothing to do with other historical states and actions. The MDP which adopts the discount criterion is denoted as $M=(S, A, P, \gamma, R)$, where $\gamma$ is the discount factor.

Reinforcement learning agents learn by interacting with external environment. Specifically, the agent observes the state $s_t \in S$ at each discrete time step $t \in [0, T]$, where T is the end time, and selects an action $a_t \in A$ used for transitioning the state $s_t \in S$ to the next state $s_{t+1} \in S$ with the probability $p$. After performing an action, the agent is usually given a scalar reward signal $r_{t+1} $, which reflects how good or bad the action was. The learning process mentioned above is repeated continuously until the agent can learn an optimal strategy, which is a mapping from the state space $S$ to the action set $A$.  

Q-learning proposed by Watkins et al. (\citeyear{watkins1992q}) is an offline reinforcement learning algorithm, and is described in \textbf{Algorithm 1}. The iteration of the Q-value function and the strategy selection are independent of each other. The approximation goal of Q-learning can be defined as $r+\gamma \max _{a^{\prime}} Q\left(s^{\prime}, a^{\prime}\right)$. The agent can choose actions according to the greedy algorithm or other non-optimal strategies.
\begin{algorithm}[t]
\caption{The Q-Table learning algorithm} 
\begin{algorithmic}[1]
\State Randomly initialize the Q table;
\For{$episode = 1,M$}
    \State Randomly initialize the $s$ state;
    \For{$step = 1,T$}
        \State Select an action $a$ according to the Q table;
        \State Execute action $a$, receive reward $r$, enter state $s^{\prime}$;
        \State {$Q(s, a)=$}
        \Statex \qquad \quad{$Q(s, a)+\alpha(r+\gamma \max _{a^{\prime} \in A} Q(s^{\prime},a^{\prime})-Q(s, a))$;}
        \State $s {\leftarrow} s^{\prime} ;$
\EndFor
\State \textbf{end for}
\EndFor
\State \textbf{end for}
\end{algorithmic}
\end{algorithm}

\section{Methods}
\subsection{Problem modeling}

The two-level quantum system (D'Alessandro and Dahleh \citeyear{d2001optimal}) is representative in filed of quantum system control. The spin 1/2 system is one of the typical two-level quantum systems for theoretical and practical research. The state $|\psi\rangle$ of the spin 1/2 system can be defined as:\begin{equation}  |\psi\rangle=\cos \frac{\theta}{2}|0\rangle+e^{t \phi} \sin \frac{\theta}{2}|1\rangle \label{equ:5}, \end{equation} where $\theta \in[0, \pi]$ and $\phi \in[0,2 \pi]$ represent the polar and phase angles respectively. A point $\vec{a}$ on the unit sphere can be defined as \begin{equation}\vec{a}=(x, y, z)=(\sin \theta \cos \phi, \sin \theta \sin \phi, \cos \theta).\end{equation}The aim is to design the control of two-level quantum system based on reinforcement learning. In the following, the problem of quantum system control based on reinforcement learning is modeled and described.

The agent in reinforcement learning learns through continuous interaction with the environment. Specific to the quantum system environment, our method divides the state space of the quantum system into a finite discrete set of states $S$. Set $A=\left\{u_{j}, j=1,2, \ldots, m\right\}$ is defined as a limited set of executable actions (unitary operations) in a quantum environment. Specifically, for the three-switch control, the $m$ is set to 3. Whenever the agent performs action $a$ and the state is transformed from $s$ to $s^{\prime}$, it will receive the feedback value, and using the fidelity as the reward: \begin{equation}
r=\left\{\begin{array}{l}
10,\ fidelity \leq 0.5\\
100,\ 0.5 < fidelity \leq 0.7\\
10000,\ fidelity > 0.7
\end{array} \label{equ:7}.\right.
\end{equation}The goal of reinforcement learning is to obtain an optimal method $\pi^*$ and the global optimal control sequence $u^*$ as Eq. (\ref{equ:4}).

For quantum systems, the agent of reinforcement learning obtains the optimal method by maximizing the long-term cumulative reward in the process of interacting with the environment of quantum systems. Therefore, the agent also needs to constantly interact with the external environment and learns through trial and error. Specifically, the permitted controls at each control step for any quantum state are $U_1$ (no control), $U_2$ (positive impulse control), and $U_3$ (negative impulse control), which is defined as follows:
\begin{equation}
\begin{aligned}
&U_{1}=e^{-\iota I_{z} \frac{\pi}{15}}, \\
&U_{2}=e^{-\iota \left(I_{z}+0.5 I_{x}\right) \frac{\pi}{15}}, \\
&U_{3}=e^{-\iota \left(I_{z}-0.5 I_{x}\right) \frac{\pi}{15}},
\end{aligned}
\end{equation}where $
I_{z}=\frac{1}{2}\left(\begin{array}{cc}
1 & 0 \\
0 & -1
\end{array}\right), I_{x}=\frac{1}{2}\left(\begin{array}{ll}
0 & 1 \\
1 & 0
\end{array}\right)$ .
The state of the quantum system in evolutionary control will be limited by the three-switch control. The agent of reinforcement learning will learn under the norms of the three-switch control in interactive learning with the environment of the quantum system. It is mainly embodied in the action selection of the agent in any quantum system state. Under the three-switch control, each action can be performed by the agent is $U_1$, $U_2$ and $U_3$.

Under the above control conditions, a global optimal control method is obtained by using proposed reinforcement learning algorithm to minimize the number of control sequences, so that the spin 1/2 system can reach the target state from the initial state.

\subsection{Enhanced reinforcement learning}

In order to improve the learning efficiency of Q learning algorithm (Watkins and Dayan \citeyear{watkins1992q}) without prior knowledge, it is important to improve the foresight ability of the learning agent. But it brings the following two problems in practice: 1) The state space increases, causing the ``dimensionality disaster", which greatly reduces the learning efficiency; 2) The visible space of the learning agent is reduced, making the agent's search process more blind.

\begin{figure*}
\begin{center}
  \includegraphics[width=0.7\textwidth]{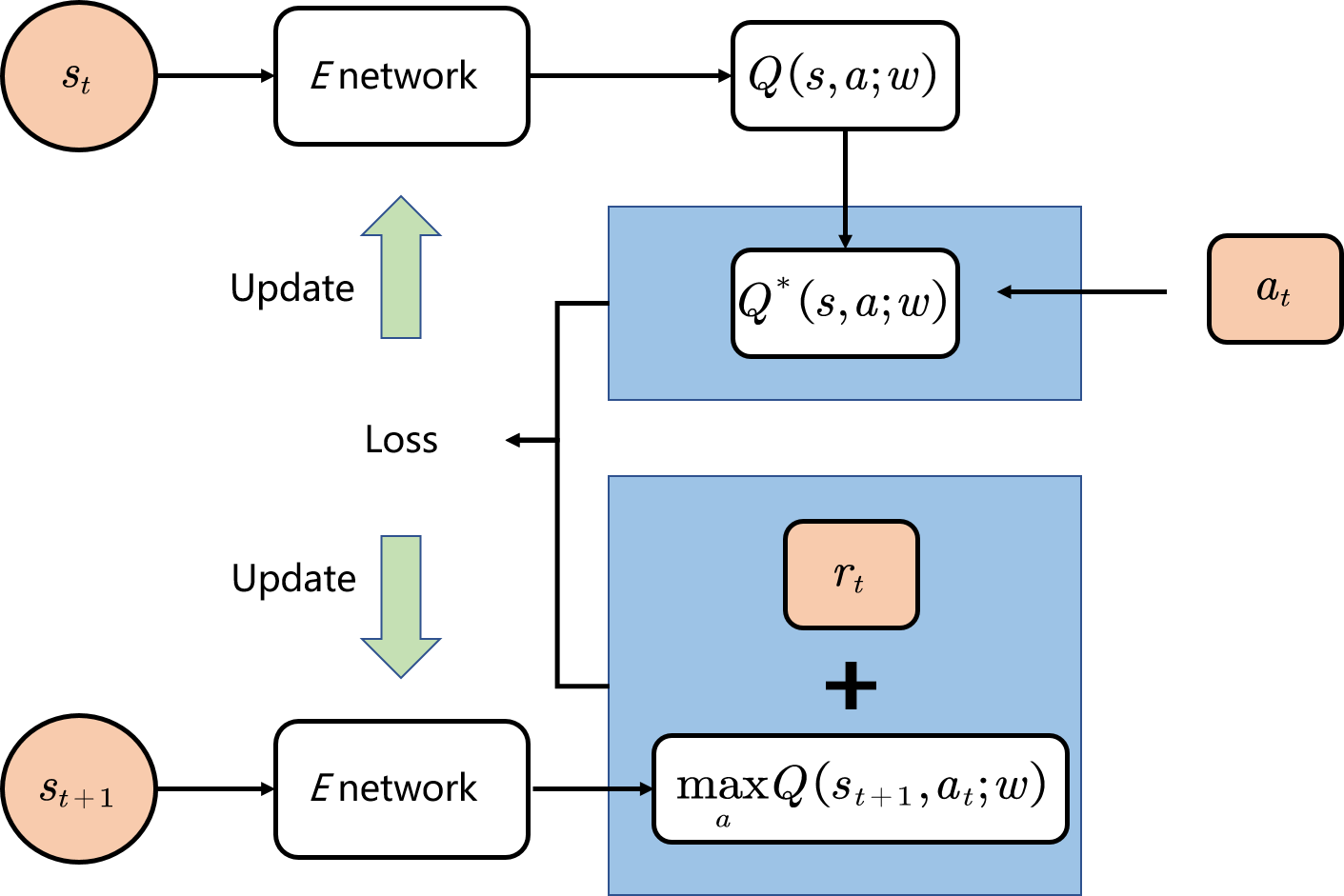}
\end{center}
\caption{An overview of enhanced reinforcement learning: the orange rectangle is given by the environment. $S_{t}$ and $S_{t+1}$ are input into the enhanced neural network which is abbreviated to E network. The algorithm selects $Q^*_{t}$ according to $a_{t}$ from $Q_{S_{t}}$and $maxQ_{t+1}$ from $Q_{S_{t+1}}$ respectively. Then calculating the loss for updating the enhanced neural network between ``the blue rectangles".}
\label{fig:1}       
\end{figure*}

To solve the above two problems, a new enhanced reinforcement learning algorithm is proposed. The enhanced reinforcement learning shown in Fig. \ref{fig:1} consists of a quantitative $Q$ table and a qualitative $V$ value heuristic function obtained by enhanced neural network.  And the description of the algorithm is shown in \textbf{Algorithm 2}. As the action $a$ is executed, the reward $r$ is obtained, state $s$ and state $s^{\prime}$ will change accordingly. The agent trains a enhanced neural network for learning a table space, which can gradually form a heuristic function to guide the agent to efficiently obtain optimal strategies for the evolution of quantum states.

\begin{algorithm}[t]
\caption{The enhanced reinforcement learning algorithm}
\begin{algorithmic}[1]
\State Initialize the Q table randomly;
\State Initialize the neural network as a qualitative layer;
\For{episode = 1,M}
\State Initialize parameters $s, \varepsilon_{0}, \gamma, \lambda, \alpha, \beta, e=0$;
\For{step = 1,T}
\State Generate $\varepsilon \in[0,1)$ randomly;
\If{$\varepsilon \leq 1-\varepsilon_{0}$} 
\State Select action $a={\arg \max} Q(s, a), {a \in A};$
\State $e=\gamma \lambda e+\frac{\partial}{\partial w} V_{\mathrm{NN}}(s) $
\Else
\State Select action $a$ randomly;
\If{$a==\arg \max _{b \in A} Q(s, b)$}
\State $e=\gamma \lambda e+\frac{\partial}{\partial w} V_{\mathrm{NN}}(s) $
\Else
\State $e=0;$
\EndIf
\State \textbf{end}
\EndIf
\State \textbf{end}
\State Execute action $a$, get reward $r$, the next state $s^{\prime}$;
\State Get $V_{\mathrm{NN}}(s)$ and $V_{\mathrm{NN}}\left(s^{\prime}\right)$ from neural network;
\State Update the enhanced neural network: 
\Statex \qquad \quad $w=w+\beta(r(s, a)+\gamma V_{\mathrm{NN}}(s^{\prime})-V_{\mathrm{NN}}(s))e; $
\State $F\left(s, a, s^{\prime}\right) =\gamma V_{\mathrm{NN }}\left(s^{\prime}\right)-V_{\mathrm{NN}}(s); $
\State $ Q(s, a)=Q(s, a)+\alpha(r (s, a)+$ 
\Statex \qquad \quad $F(s, a, s^{\prime})+\gamma \max Q(s^{\prime}, a^{\prime})-Q(s, a))$
\State $s \leftarrow s^{\prime} ;$
\EndFor
\State \textbf{end for}
\EndFor
\State \textbf{end for}
\end{algorithmic}
\end{algorithm}

\subsubsection{Heuristic function based on enhanced neural network}
\begin{figure}
\centering
  \includegraphics[width=0.7\linewidth]{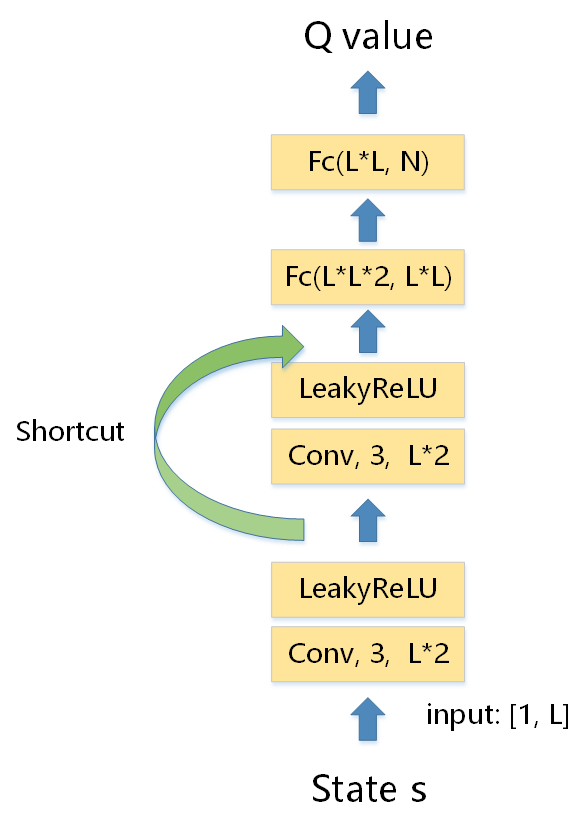}
\caption{The enhanced neural network architecture: For each state s fed into the network, the network extracts features and outputs Q values.}
\label{fig:2}
\end{figure}

To build a generalization and foresight capacity to avoid the blind behavior of the agent, a heuristic function based on enhanced neural network is proposed. The Q-table in enhanced reinforcement learning is updated with the execution of actions. At the same time, the enhanced neural network shown in Fig. \ref{fig:2} is trained, and a $V$ value fitting surface is gradually developed. The heuristic function based on enhanced neural network thereby shows up to guide the optimization and updating of the new quantum system control method based on enhanced reinforcement learning (QSC-ERL). Inspired by common convolutional neural network (Gu et al. \citeyear{gu2018recent}) and residual neural network (He et al. \citeyear{he2016deep}), the enhanced neural network can make full use of the extracted features. The state $s$ is the input of the neural network, and the Q values got by the probability of actions is the output, where $N$ is the number of actions. In order to obtain the nonlinear characteristics more comprehensively, the Leaky ReLU is selected as activation function to give all negative values a non-zero slope.
The heuristic function $F\left(s, a, s^{\prime }\right)$ participating in the update of the Q table takes $s$ and $s^{\prime}$ as input and gets the $V$ value output which is defined as $V_{N N}(s)$ and $V_{N N}\left(s^{\prime}\right)$ in state $s$ and $s^{\prime}$ respectively. And the heuristic function is defined as
\begin{equation}
F\left(s, a, s^{\prime}\right)=\gamma V_{N N}\left(s^{\prime}\right)-V_{N N}(s).
\end{equation}

\subsubsection{The parameters updating method}
To build an effective parameters updating method, and accelerate training speed of QSC-ERL, the eligibility trace (Singh and Suttun \citeyear{singh1996reinforcement}) is introduced. The error obtained by updating can be passed back several steps to speed up the learning of the enhanced neural network and provide an effective inspiration for the whole algorithm.

The learning of Q table can be defined as 
\begin{equation}
\begin{aligned}
Q(s, a)=&Q(s, a)+\alpha[r(s, a, s^{\prime})+F(s, a, s^{\prime})\\
&+\gamma \max _{a^{\prime}} Q(s^{\prime}, a^{\prime})-Q(s, a)],\label{equ:10}
\end{aligned}
\end{equation}and the updating of $V$ values can be defined as 
\begin{equation}
V(s)=\max _{a} Q(s, a).\label{equ:8}
\end{equation}
When the agent performs a non-greedy action, its next state $s^{\prime}$ often does not obtain the largest Q value. The QSC-ERL will update the current state-action paired Q value according to the $V$ value of the next state obtained by the greedy strategy. For updating the enhanced neural network, when the agent requires the $V$ value according to the greedy strategy, the eligibility trace is also updated. When the agent performs a non-greedy strategy, the eligibility trace is set as 0, preventing the error from propagating backward.

To update the weights of the enhanced neural network, the gradient descent method is adopted which can be defined as 
\begin{equation}
\begin{aligned}
\Delta w_{t}=&\beta(r(s_{t})+\gamma V_{\mathrm{NN}}(s_{t+1})-V_{\mathrm{NN}}(s_{t})) \times \\
&\sum_{k=0}^{t}(\gamma \lambda)^{tk} \frac{\partial}{\partial w} V_{\mathrm{NN}}(s_{k}), \label{equ:12}
\end{aligned}
\end{equation}
where $\beta$ is the learning rate, $0<\beta<1$, $\lambda$ is the eligibility trace coefficient, $0<\lambda<1$.

The agent updates the weight of the neural network through the difference value $r\left(s_{t}\right)+\gamma V_{\mathrm{NN}}\left(s_{t +1}\right)-V_{\mathrm{NN}}\left(s_{t}\right)$ between the next predicted $V$ value of state $s$ and the current target $V$ value. The difference value can be used for updating the $V$ value in other state. If the eligibility trace is defined as
\begin{equation}
e_{t}=\sum_{k=0}^{t} \gamma \lambda \frac{\partial}{\partial w} V_{\mathrm{NN}}(s_{ k})=\gamma \lambda e_{t-1}+\frac{\partial}{\partial w} V_{\mathrm{NN}}(s_{t}),
\end{equation}
Eq. (\ref{equ:12}) can be rewritten as
\begin{equation}
\Delta w_{t}=\beta(r(s_{t})+\gamma V_{\mathrm{NN}}(s_{t+1})-V_{\mathrm{NN}}(s_{t}))e_{t}.\label{equ:14}
\end{equation} It is easy for modifying the weights from the hidden layer of the neural network to the output layer, and then modify the weights from the input layer to the hidden layer through the back propagation.

The QSC-ERL is carried out synchronously in the learning of Q-Table and the enhanced neural network. The Q-Table based reinforcement learning can obtain more accurate results, but the speed of learning is slow. The enhanced neural network is not accurate enough, but it has better generalization performance. In the initial stage of learning, the effect is not obvious. But with continuous learning, by using the parameters updating method, the enhanced neural network is gradually established the trend information, and the convergence speed can be greatly improved.

\section{Simulation experiments}
\subsection{Settings}
Since it is difficult to verify the validity and efficiency of the algorithm in real quantum computers, the realization of the experiment is inseparable from the quantum control landscapes (Chakrabarti and Rabitz \citeyear{chakrabarti2007quantum}). The simulation experiment is implemented by Python programming language and Linalg tool library. Full training for a given scenario can be achieved on a single CPU+GPU workstation (CPU: Intel Xeon Gold 5218, GPU: GeForce RTX 2080 Ti 11G).  The state space of the quantum system will be reconstructed from the initial state $s_{initial}=|\psi_{initial}\rangle$ to the target state $s_{target}=|\psi_{target}\rangle$. The state set is $S=\{s_{i}=|\psi_{i}\rangle, i=1,2, \ldots, n\}$, and the executable action set is $A=\{a_{j}=U_{j}, j=1,2, \ldots, m_{\circ}\}$. For the spin 1/2 system, the initial state is set as $|\psi_{initial}\rangle(\theta=(\pi / 60), \phi=(\pi / 30))$, and the target state is $|\psi_{target }\rangle(\theta=(41 \pi / 60), \phi=(29 \pi / 30))$. The Eq. (\ref{equ:3}) and Eq. (\ref{equ:5}) can be utilized to construct the whole quantum simulation environment. The setting of the reward in QSC-ERL is according to the Eq. (\ref{equ:7}). Here is the parameter settings shown as table \ref{tab:0}.

\begin{table}
\caption{The parameter settings of the QSC-ERL}
\label{tab:0}       
\begin{tabular}{ll}
\hline\noalign{\smallskip}
 Name & Value  \\
\noalign{\smallskip}\hline\noalign{\smallskip}
maximum episode & 500 \\
learning rate  & 0.01 \\
reward decay & 0.9  \\
e greedy &   0.99    \\
memory size&   2000     \\
\noalign{\smallskip}\hline
\end{tabular}
\end{table}

\subsection{Evaluation index}
Fidelity is a evaluation index to measure the distance between density operators. It allows us to compare how the state of the system at any given moment is different from the initial state, or how the state of a system is different from a reference state. It allows us to measure quantitatively how different two states really are. For two density matrices $\rho$, $\sigma$ it is generalized as the largest fidelity between any two purifications of the given states. And the fidelity function can be defined as
\begin{equation}
F(\rho, \sigma)=(tr\sqrt{\sqrt{\rho} \sigma \sqrt{\rho}})^{2},
\end{equation}where $\rho$ and $\sigma$ are the density matrix of source information and target information respectively.

\subsection{Results and analysis}

The simulation experiments is carried out under the three switch control paradigm. The goal of the experiment is to control the spin 1/2 system from the initial state $|\psi_{initial}\rangle$ to the target state $|\psi_{target}\rangle$. The main purpose is to explore the effectiveness of reinforcement learning algorithm for solving quantum control problem.

Therefore, the simulation experiments is divided for two parts: one is that the tabular Q-learning (TQL) (Sutton and Barto \citeyear{sutton2018reinforcement}), deep Q-learning (DQL) (Mnih et al. \citeyear{mnih2015human}) and policy gradient (PG) (Sutton et al. \citeyear{sutton2000policy}) are applied to explore the effectiveness of reinforcement learning algorithm for solving quantum control problem. The other is that the NN-QSC (Fosel et al. \citeyear{fosel2018reinforcement}) and the DRL-QSC (An and Zhou \citeyear{an2019deep}) are compared for verifying that the proposed QSC-ERL performed better than its peers. The parameters of the reinforcement learning algorithms involved in the experiment are set as follows: For all state-action paired, the Q value is initialized to 0, the discount factor is $\gamma=0.99$, the learning rate is $\alpha=0.1$, and the action selection probability is initialized to $1/3$. 

\begin{figure*}
  \includegraphics[width=1\textwidth]{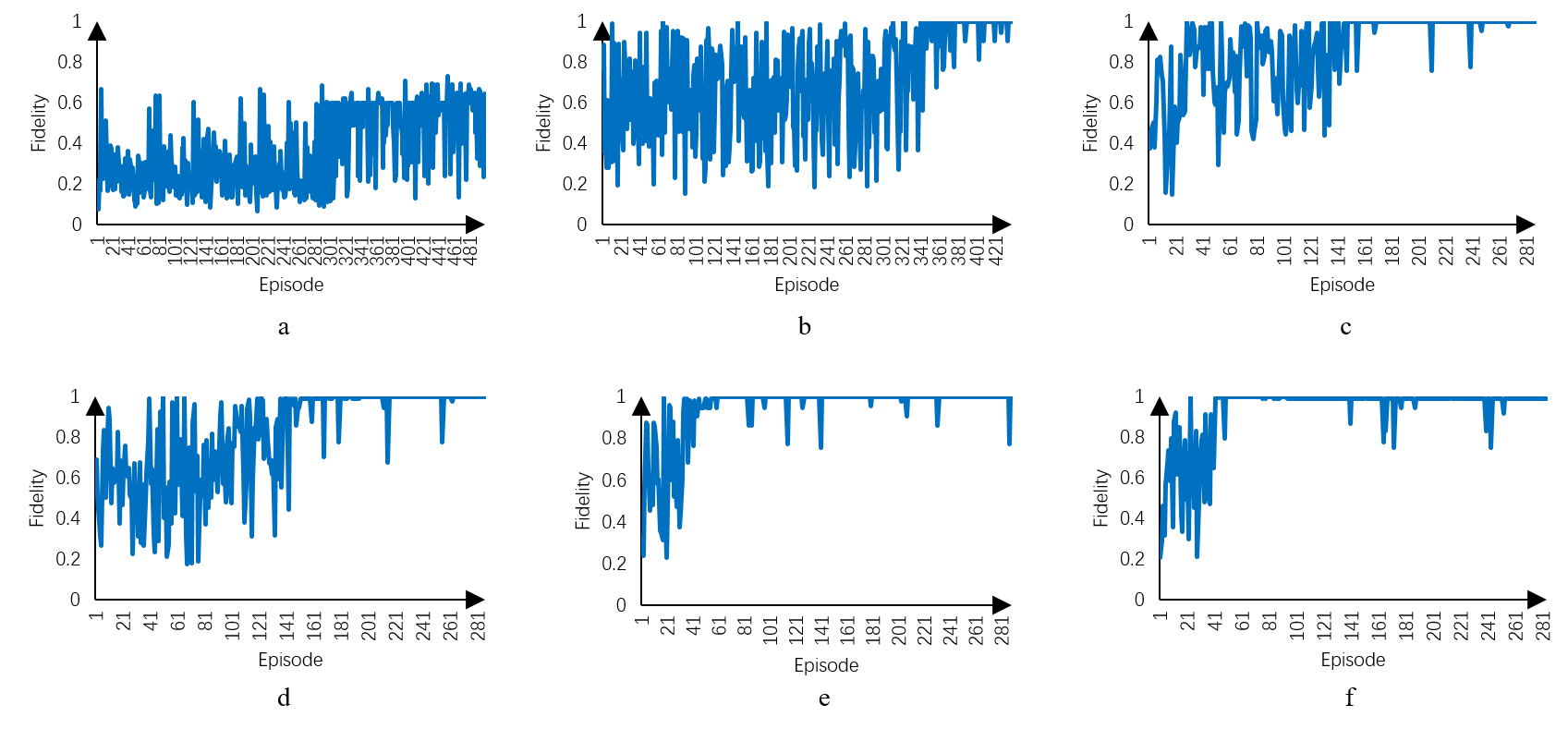}
\caption{The comparison of fidelity between algorithms. (a)Fidelity of the TQL algorithm. (b) Fidelity of the PG algorithm. (c) Fidelity of the DQL algorithm. (d) Fidelity of the NN-QSC algorithm. (e) Fidelity of the DRL-QSC algorithm. (f) Fidelity of the QSC-ERL algorithm.}
\label{fig:3}       
\end{figure*}
Fig. \ref{fig:3} shows the comparison of fidelity between algorithms, where the X-axis is the number of episode, and the Y-axis is fidelity. It can be seen from Fig. \ref{fig:3} that reinforcement learning has certain effects on solving the quantum system control problems. Since the TQL algorithm can not converge rapidly during training, the fidelity is the lowest. The PG algorithm has better convergence, but only a little bit at one episode. The other four methods have good results. The DQL algorithm adopted convolutional neural network to guide the learning of the Q learning algorithm. Although it solves the problem of not being able to update Q table well when there are many actions, it is difficult for simple neural networks to learn the useful features of quantum systems. If want to get a high fidelity between the final state and the target state, it needs to train with more episodes and data set. Due to the high correlation between states in the training process, the NN-QSC and the DRL-QSC may fall into local optimum or be difficult to converge. Our QSC-ERL use the enhanced neural network to effectively make use of the differential features before and after the evolution of the quantum state. By introducing the eligibility trace to update parameters, The QSC-ERL can quickly find the optimal control strategy of the quantum system. Table \ref{tab:1} shows the number of episodes when the fidelity can get the maximum, and total number of episodes is set to 500. The data is taken from the average value of 100 experiments. It represents that the ability of algorithms can make the quantum system from the initial state to the desired target state.
\begin{table}
\caption{The comparison of the number of episodes between algorithms}
\label{tab:1}       
\begin{tabular}{lll}
\hline\noalign{\smallskip}
 Name & Episodes & Fidelity  \\
\noalign{\smallskip}\hline\noalign{\smallskip}
TQL& 452  & 0.73\\
PG& 311 & 0.99 \\
DQL& 135 & 0.99 \\
NN-QSC&   171   &    0.99  \\
DRL-QSC&   60   &   0.99   \\
QSC-ERL &    42  &    0.99 \\
\noalign{\smallskip}\hline
\end{tabular}
\end{table}
The experimental results show that most methods converge after training and make the quantum system from the initial state to the desired target state. Specifically, for the TQL, the maximum of the fidelity is 0.73, and others can reach 0.99 after total training. The PG requires about 311 episodes and the DQL requires about 135 episodes. It means that the reinforcement learning based on neural network has the better performance in some degree than the common RL algorithm. The NN-QSC requires about 171 episodes to control the evolution of the quantum system from the initial state to the target state while the DRL-QSC requires about 60 episodes, and the QSC-ERL requires the 42 episodes. So our proposed QSC-ERL algorithm is faster than the NN-QSC and the DRL-QSC for controlling the evolution of the quantum system from the initial state to the target state.

\section{Conclusion}
In this paper, a quantum system control method based on enhanced reinforcement learning (QSC-ERL) is proposed to achieve the learning control of the spin 1/2 system. A satisfactory control strategy is obtained through enhanced reinforcement learning so that the quantum system can be evolved accurately from the initial state to the target state. Compared with other methods, our method can achieve the quantum system control with high fidelity, and improve the control efficiency of quantum systems.

It should be noted that our method is sufficient for the evolution of quantum state in spin 1/2 system. Other difficult quantum control problems include quantum error correction based on bosonic codes (Michael et al. \citeyear{michael2016new}) and quantum state preparation in the single-photon manifold (Vrajitoarea et al. \citeyear{vrajitoarea2020quantum}). And it is a valuable work to conduct a study on providing solutions by using learning theories (Li et al. \citeyear{li2018visual}; Zhang and Wang \citeyear{zhang2020hybrid}) and neural network (Xu et al. \citeyear{xu2019investigation}; Hu et al. \citeyear{hu2020multiple}), which is also one of our next research.

\begin{acknowledgements}
The authors would like to thank the anonymous reviewers and editors for their comments that improved the quality of this paper. This work is supported by the National Natural Science Foundation of China (62071240, 61802175), the Natural Science Foundation of Jiangsu Province (BK20171458), and the Priority Academic Program Development of Jiangsu Higher Education Institutions (PAPD).
\end{acknowledgements}

\section*{Declarations}

\textbf{Conflict of interest} The authors declare that they have no conflict of interest.\\
\textbf{Ethical statement} Articles do not rely on clinical trials.\\
\textbf{Human and animal participants} All submitted m-\\anuscripts containing research which does not involve human participants and/or animal
experimentation.

%
%



\end{document}